\documentclass[aps,preprint,nofootinbib]{revtex4}
\usepackage{graphics}
\usepackage{slashed}
\usepackage{amssymb}
\usepackage{lscape}
\usepackage{amsmath}
\usepackage{rotating}
\usepackage{bm}
\usepackage{graphicx}
\graphicspath{ {images/} }
\begin{document}
\title{Gravitons emission during pre-inflation from unified spinor fields}
\author{$^{2}$ Luis Santiago Ridao\footnote{E-mail address: santiagoridao@hotmail.com}, $^{2}$ Marcos R. A. Arcod\'{\i}a\footnote{E-mail address: marcodia@mdp.edu.ar},  $^{1,2}$ Mauricio Bellini
\footnote{E-mail address: mbellini@mdp.edu.ar} }
\address{$^1$ Departamento de F\'isica, Facultad de Ciencias Exactas y
Naturales, Universidad Nacional de Mar del Plata, Funes 3350, C.P.
7600, Mar del Plata, Argentina.\\
$^2$ Instituto de Investigaciones F\'{\i}sicas de Mar del Plata (IFIMAR), \\
Consejo Nacional de Investigaciones Cient\'ificas y T\'ecnicas
(CONICET), Mar del Plata, Argentina.}
\begin{abstract}
We obtain the equation that describe the conditions of quantization for neutral massless bosons on an arbitrary curved space-time, obtained using a particular theoretical formalism developed in a previous work\cite{MM}. In particular, we study the emission of neutral massless $(1, 2)\hbar$-spin bosons during pre-inflation using the recently introduced unified spinor field theory. We conclude that during pre-inflation (which is governed by an vacuum equation of state), is emitted gravitational radiation, which could be detected in the future, as primordial gravitational radiation.
\end{abstract}
\maketitle

\section{INTRODUCTION}

It is well known that Heisenberg suggested an unified quantum field theory of a fundamental spinor field describing all matter fields in their interactions\cite{hei1,hei2}. In his theory the masses and interactions of particles are a consequence of a self-interaction term of the elementary spinor field. The fact that manifolds with no-Euclidean geometry can help uncover new features of quantum matter makes it desirable to create manifolds of controllable shape and to develop the capability to add in synthetic gauge fields\cite{hohuang}.

On the other hand, in a previous work we have developed construct a pure geometric spinor field theory on an arbitrary curved background, which is considered a Riemannian manifold. In the theory the spinor field $\hat{\Psi}^{\alpha}$ is responsible for the displacement of the extended Weylian manifold\cite{weyl} with respect to the Riemannian background and the covariant derivative of the metric tensor in the Riemannian background manifold is null\footnote{We denote with a $\nabla$, the Riemannian-covariant derivative and with a $\Delta$ the Riemannian variation of some arbitrary tensor: $\Delta g_{\alpha\beta}=\nabla_{\gamma} g_{\alpha\beta} \,dx^{\gamma}=0$.}. However, the Weylian covariant derivative on the extended Weylian manifold\footnote{We denote the covariant derivative on the extended Weylian manifold with a $ \|  $.}, is nonzero: $ g^{\alpha\beta}_{\quad \|\gamma} \neq 0$. In this formalism they are considered the couplings of the spinor fields with the background and their self-interactions in a generic manner. The theory is worked in 8 dimensions, 4 of them related to the space-time coordinates ($x^{\mu}$), and the other 4 related to the inner space ($\phi^{\mu}$), described by compact coordinates. The former have the spin components as canonical momentums: ($s_{\mu}$).

The letter is organized as follows: In Sect. \ref{ei} we have described the space-time structure from a quantum approach to recover a line element on a background Riemannian manifold.
In Sect. \ref{spi} we have exposed the spinor field formalism for bosons and the dynamic equations. In Sect. \ref{nb} we have studied the particular case of the neutral bosons on arbitrary curved backgrounds. In Sect. \ref{pin} we studied the example of neutral massless bosons in the pre-inflationary epoch. In particular, we consider the the emission of massless spinor fields with spins $s=\hbar$ and $s=2\hbar$. Finally, in Sect. \ref{fc} we develop some final comments.

\section{EINSTEIN-HILBERT ACTION AND QUANTUM STRUCTURE OF SPACE TIME}\label{ei}

If we deal with an orthogonal basis,
the curvature tensor will be written in terms of the connections:
$R^{\alpha}_{\,\,\,\beta\gamma\delta} = \Gamma^{\alpha}_{\,\,\,\beta\delta,\gamma} -  \Gamma^{\alpha}_{\,\,\,\beta\gamma,\delta}
+ \Gamma^{\epsilon}_{\,\,\,\beta\delta} \Gamma^{\alpha}_{\,\,\,\epsilon\gamma} - \Gamma^{\epsilon}_{\,\,\,\beta\gamma}
\Gamma^{\alpha}_{\,\,\,\epsilon\delta}$). The Einstein-Hilbert (EH) action for an arbitrary matter lagrangian density ${\cal L}$
\begin{equation}
{\cal I} = \int d^4 x \sqrt{-g} \left[ \frac{R}{2\kappa}+ {\cal L} \right],
\end{equation}
after variation, is given by
\begin{equation}\label{delta}
\delta {\cal I} = \int d^4 x \sqrt{-g} \left[ \delta g^{\alpha\beta} \left( G_{\alpha\beta} + \kappa T_{\alpha\beta}\right)
+ g^{\alpha\beta} \delta R_{\alpha\beta} \right],
\end{equation}
where $\kappa = 8 \pi G$, $G$ is the gravitational constant, $g^{\alpha\beta} \delta R_{\alpha\beta} =\delta\Theta(x^{\alpha})$, such that 
$\delta\Theta(x^{\alpha})$ is an arbitrary scalar field, and $T_{\alpha\beta}$ is the energy-momentum tensor defined by
\begin{equation}
T_{\alpha\beta}= 2 \frac{\delta{\cal L}}{\delta g^{\alpha\beta}} - g_{\alpha\beta} {\cal L}.
\end{equation}
When the flux $\delta\Theta(x^{\alpha})$ that cross the Gaussian-like hypersurface defined on an arbitrary region of the spacetime,
is zero, the resulting equations that minimize the EH action, are the background Einstein equations: $G_{\alpha\beta} + \kappa\, T_{\alpha\beta}=0$. However, when this flux is nonzero,
one obtains in the last term of the eq. (\ref{delta}). This flux becomes zero when there are no sources within this hypersurface. Hence, in order to make $\delta {\cal I}=0$ in Equation (\ref{delta}), we must consider the condition: $
G_{\alpha\beta} + \kappa T_{\alpha\beta} = \Lambda\,
g_{\alpha\beta}$, where $\Lambda$ is the cosmological constant. On the other hand, we can make the transformation
\begin{equation}\label{ein}
\bar{G}_{\alpha\beta} = {G}_{\alpha\beta} - \Lambda\, g_{\alpha\beta},
\end{equation}
where the scalar field $\delta\Theta$ complies $\Box \delta\Theta =0$\cite{rb}, and the transformed Einstein equations with the equation of motion for the transformed gravitational waves, hold
\begin{equation}
\bar{G}_{\alpha\beta} = - \kappa\, {T}_{\alpha\beta}. \label{e1} \\
\end{equation}
The Equation (\ref{e1}) give us the Einstein equations with cosmological
constant included. Notice that the scalar field $\delta\Theta(x^{\alpha})$ appears
as a scalar flux of some 4-vector with components $\delta W^{\alpha}$: 
\begin{equation}\label{uchi}
\left[\delta W^{\alpha}\right]_{||\alpha} = \delta\Theta(x^{\alpha}),
\end{equation}
through the closed hypersurface $\partial{\cal M}$, which is situated in any region of space-time. Here,
$\delta W^{\alpha}=\delta\Gamma^{\epsilon}_{\beta\epsilon} g^{\beta\alpha}-\delta \Gamma^{\alpha}_{\beta\gamma} g^{\beta\gamma}$\footnote{We define the covariant derivative of some vector field $\Upsilon^{\beta}$: $\left[\Upsilon^{\beta}\right]_{||\alpha}$
\begin{equation}
\left[\Upsilon^{\beta}\right]_{||\alpha} = \nabla_{\alpha}\Upsilon^{\beta} + \xi^2 \,\delta\Gamma^{\beta}_{\epsilon\alpha}\Upsilon^{\epsilon},
\end{equation}
where $\xi$ is the self-interaction constant, $\nabla_{\alpha}\Upsilon^{\beta}$ is the covariant derivative on the Riemann manifold and $\delta\Gamma^{\beta}_{\epsilon\alpha}$ is the displacement of the manifold with respect to the Riemann one.}. In this work we shall use a recently introduced extended Weylian manifold\cite{MM} to describe quantum geometric spinor fields
$\hat{\Psi}^{\alpha}$, where the connections are
\begin{equation}\label{ga}
\hat{\Gamma}^{\alpha}_{\beta\gamma} = \left\{ \begin{array}{cc}  \alpha \, \\ \beta \, \gamma  \end{array} \right\}+ \hat{\Psi}^{\alpha}\,g_{\beta\gamma}.
\end{equation}
Here
\begin{equation}\label{uch}
\hat{\delta{\Gamma}}^{\alpha}_{\beta\gamma}=\hat{\Psi}^{\alpha}\,g_{\beta\gamma},
\end{equation}
describes the quantum displacement of the extended Weylian manifold with respect to the classical Riemannian background, which is described by the Levi-Civita symbols in (\ref{ga}), and the variation of the Ricci tensor is
\begin{equation}
\hat{\delta{R}}_{\beta\gamma} = \left(\hat{\delta\Gamma}^{\alpha}_{\beta\alpha} \right)_{\| \gamma} - \left(\hat{\delta\Gamma}^{\alpha}_{\beta\gamma} \right)_{\| \alpha},
\end{equation}
where $\hat{\delta\Gamma}^{\alpha}_{\beta\alpha}=\hat{\Psi}^{\alpha}\,g_{\beta\gamma}$.

\subsection{Quantum structure of space-time}

In order to describe the quantum structure of space time we consider a the variation $\delta\hat{X}^{\mu}$ of the quantum operator $\hat{X}^{\mu}$:
 \begin{displaymath}
\hat{X}^{\alpha}(x^{\nu}) = \frac{1}{(2\pi)^{3/2}} \int d^3 k \, \hat{\gamma}^{\alpha} \left[ b_k \, \hat{X}_k(x^{\nu}) + b^{\dagger}_k \, \hat{X}^*_k(x^{\nu})\right],
\end{displaymath}
such that $b^{\dagger}_k$ and $b_k$ are the creation and destruction operators of space-time, such that $\left< B \left| \left[b_k,b^{\dagger}_{k'}\right]\right| B  \right> = \delta^{(3)}(\vec{k}-\vec{k'})$and $\hat{\gamma}^{\alpha}$ are $4\times 4$-matrices that comply with the Clifford algebra. Moreover, we shall define in the analogous manner the variation $\delta\hat{\Phi}^{\mu}$ of the quantum operator $\hat{\Phi}^{\mu}$ that describes the quantum inner space:
\begin{displaymath}
\hat{\Phi}^{\alpha}(\phi^{\nu}) = \frac{1}{(2\pi)^{3/2}} \int d^3 s \, \hat{\gamma}^{\alpha} \,\left[ c_s \, \hat{\Phi}_s(\phi^{\nu}) + c^{\dagger}_s \, \hat{\Phi}^*_s(\phi^{\nu})\right],
\end{displaymath}
such that $c^{\dagger}_s$ and $c_s$ are the creation and destruction operators of the inner space, such that $\left< B \left| \left[c_s,c^{\dagger}_{s'}\right]\right| B  \right> = \delta^{(3)}(\vec{s}-\vec{s'})$. In our case the background quantum state can be represented in a ordinary Fock space in contrast with LQG\cite{aa}, where operators are qualitatively different
from the standard quantization of gauge fields. These operators can be applied to some background  quantum state, and describes a Fock space on an arbitrary Riemannian curved space time $\left|B\right>$, such that they comply with
\begin{equation}\label{dif}
\delta\hat{X}^{\mu}\left|B\right> = dx^{\mu}\left|B\right>, \qquad \delta\hat{\Phi}^{\mu}\left|B\right> = d\phi^{\mu}\left|B\right>.
\end{equation}
The states $\left|B\right>$ do not evolves with time because we shall consider the Heisenberg representation, in which only the operators evolve with time so that the background expectation value of the manifold displacement is null: $\left<B\left|\hat{\delta\Gamma}^{\alpha}_{\beta\gamma}\right|B\right>=0$. In order to describe the effective background space-time, we shall consider the line element
\begin{equation}\label{line}
dl^2 \delta_{BB'}= dx^2 \delta_{BB'} + d\phi^2 \delta_{BB'} = \left<B\right| \hat{\delta X}_{\mu} \hat{\delta X}^{\mu} \left| B'\right> + \left<B\right| \hat{\delta \Phi}_{\mu} \hat{\delta \Phi}^{\mu} \left| B'\right>,
\end{equation}
where $\phi^{\alpha}$ are the four compact dimensions related to their canonical momentum components $s^{\alpha}$ that describe the spin. The variations and differentials of the operators $\hat{X}^{\mu}$ and $\hat{\Phi}^{\mu}$ on the extended Weylian manifold, are given respectively by
\begin{eqnarray}
\delta\hat{X}^{\mu}\left| B\right> &=& \left(\hat{X}^{\mu}\right)_{\|\alpha} dx^{\alpha}\left| B\right>, \qquad \delta\hat{\Phi}^{\mu} \left| B\right>= \left(\hat{\Phi}^{\mu}\right)_{\|\alpha} d\phi^{\alpha}\left| B\right>, \\
d\hat{X}^{\mu} \left| B\right>&=& \left(\hat{X}^{\mu}\right)_{,\alpha} dx^{\alpha}\left| B\right>, \qquad d\hat{\Phi}^{\mu} \left| B\right>= \left(\hat{\Phi}^{\mu}\right)_{,\alpha} d\phi^{\alpha}\left| B\right>,
\end{eqnarray}
with covariant derivatives
\begin{eqnarray}
\left(\hat{X}^{\mu}\right)_{\|\beta}\left| B\right> &=& \left[\nabla_{\beta} \hat{X}^{\mu} + \hat{\Psi}^{\mu} \hat{X}_{\beta} - \hat{X}^{\mu} \hat{\Psi}_{\beta}\right]\left| B\right>, \\
\left(\hat{\Phi}^{\mu}\right)_{\|\beta}\left| B\right> &=& \left[\nabla_{\beta} \hat{\Phi}^{\mu} + \hat{\Psi}^{\mu} \hat{\Phi}_{\beta} - \hat{\Phi}^{\mu} \hat{\Psi}_{\beta}\right]\left| B\right>.
\end{eqnarray}

\subsection{Bi-vectorial structure of inner space}

We shall consider the squared of the $\hat{\delta\Phi}$-norm on the bi-vectorial space, and the squared $\hat{\delta X}$-norm on the vectorial space, are
\begin{eqnarray}
\underleftrightarrow{\delta\Phi} \overleftrightarrow{\delta\Phi} &\equiv& \left( \hat{\delta\Phi}_{\mu} \hat{\delta\Phi}_{\nu} \right) \left( \bar{\gamma}^{\mu} \bar{\gamma}^{\nu}\right), \\
\underrightarrow{\delta{X}} \overrightarrow{\delta{X}} & \equiv &  \hat{\delta{X}}_{\alpha} \hat{\delta{X}}^{\alpha}.
\end{eqnarray}
such that $\hat{\Phi}^{\alpha}=\phi \,\bar{\gamma}^{\alpha} $ and $\hat{X}^{\alpha}=x\, \bar{\gamma}^{\alpha} $ are respectively the components of the inner and coordinate spaces. Furthermore,  $\bar{\gamma}_{\mu}$ are the ($4\times 4$) Dirac matrices that generate the vectorial and bi-vectorial structure of the space time
\begin{eqnarray}
\left<B\left|\hat{X}_{\mu} \hat{X}^{\mu} \right|B\right> &=& x^2\, \,\mathbb{I}_{4\times 4}, \\
\left<B\left|\left( \hat{\Phi}_{\mu} \hat{\Phi}_{\nu} \right) \left( \bar{\gamma}^{\mu} \bar{\gamma}^{\nu}\right) \right|B\right>&=&
\left<B\left|\frac{1}{4} \left\{ \hat{\Phi}_{\mu} ,\hat{\Phi}_{\nu} \right\} \left\{ \bar{\gamma}^{\mu} ,\bar{\gamma}^{\nu}\right\} -\frac{1}{4}
\left[ \hat{\Phi}_{\mu}, \hat{\Phi}_{\nu} \right] \left[ \bar{\gamma}^{\mu}, \bar{\gamma}^{\nu}\right] \right|B\right>\nonumber \\
&=&\phi^2 \,\mathbb{I}_{4\times 4}. \nonumber
\end{eqnarray}
The $\bar\gamma^{\mu}$ matrices, comply with the Clifford algebra
\begin{equation}
\bar{\gamma}^{\mu} = \frac{\bf{I}}{3!} \left(\bar{\gamma}^{\mu}\right)^2\,\epsilon^{\mu}_{\,\,\alpha\beta\nu} \bar{\gamma}^{\alpha\beta}  \bar{\gamma}^{\nu} , \qquad \left\{\bar{\gamma}^{\mu}, \bar{\gamma}^{\nu}\right\} =
2 g^{\mu\nu} \,\mathbb{I}_{4\times 4}, \nonumber
\end{equation}
where ${\bf{I}}={\gamma}^{0}{\gamma}^{1}{\gamma}^{2}{\gamma}^{3}$, $\mathbb{I}_{4\times 4}$ is the identity matrix, $\bar{\gamma}^{\alpha\beta}=\frac{1}{2} \left[\bar{\gamma}^{\alpha}, \bar{\gamma}^{\beta}\right]$. In this paper we shall consider the Weyl basis on a Minkowsky spacetime (in cartesian coordinates): $\left\{ \gamma^a,\gamma^b\right\} = 2 \eta^{ab} \mathbb{I}_{4\times 4}$
\begin{eqnarray}
&& \gamma^0= \,\left(\begin{array}{ll}  0 & \mathbb{I} \\
\mathbb{I}  &  0 \ \end{array} \right),\qquad
\gamma^1=  \left(\begin{array}{ll} 0 &  -\sigma^1 \\
\sigma^1 & 0  \end{array} \right),  \nonumber \\
&& \gamma^2= \left(\begin{array}{ll} 0 &  -\sigma^2 \\
\sigma^2 & 0  \end{array} \right),  \qquad \gamma^3= \left(\begin{array}{ll} 0 &  -\sigma^3 \\
\sigma^3 & 0  \end{array} \right),\nonumber
\end{eqnarray}
such that the Pauli matrices are
\begin{eqnarray}
&& \sigma^1 = \left(\begin{array}{ll} 0 & 1 \\
1  & 0  \end{array} \right), \quad \sigma^2 = \left(\begin{array}{ll} 0 & -i \\
i  & 0  \end{array} \right), \quad \sigma^3 = \left(\begin{array}{ll} 1 & 0 \\
0  & -1  \end{array} \right). \nonumber
\end{eqnarray}

\section{SPINOR FIELD}\label{spi}

The  expressions (\ref{uchi}) and (\ref{ga}), give us
\begin{equation}\label{ps}
\frac{\hat{\delta W}^{\alpha}}{\delta l} = 3 \hat{\Psi}^{\alpha},
\end{equation}
$l$-being the Weylian 4-length. The self-interacting effects do not necessary preserve the Riemannian flux of matter fields along the Gaussian hypersurface, so that $ \left(\delta W^{\alpha}\right)_{\|\alpha}=\nabla_{\alpha} \delta W^{\alpha}+\xi^2 \, \delta W^{\alpha} \hat{\Psi}_{\alpha}$. Notice that when the coupling constant is zero: $\xi=0$, the Riemannian flux on the extended Weylian manifold is equal to the flux on the Riemannian one. The flux equation can be rewritten using (\ref{ps}), so that
\begin{equation}\label{lig}
\nabla_{\alpha} \hat{\Psi}^{\alpha}+\xi^2 \, \hat{\Psi}^{\alpha} \hat{\Psi}_{\alpha}=\frac{1}{3}\frac{\hat{\delta\Theta}}{\delta l}.
\end{equation}
However, when we describe matter fields, the coupling $\xi$ is nonzero. In general, $\xi$ depends on the theory under study (i.e. on the group representation of the spinor fields), and can be proportional to some physical property of the field (mass, charge, etc). Spinors with $\xi=0$, do not describe matter fields, but geometric fields.

In this framework, we can define respectively the slash and vector quantum fields $\slashed{\Psi}     = \hat{\Psi}_{\alpha} \bar{\gamma}^{\alpha}$, $\overleftrightarrow{\Psi} = \Psi_{\alpha} \bar{\gamma}^{\alpha}$. The $4$-vector components are $\hat\Psi_{\alpha}=\frac{\hat{\delta\Theta}}{\hat{\delta{\Phi}}^{\alpha}}$, where the flux of $\hat{\Psi}^{\alpha}$-field through the Gaussian hypersurface in eq. (\ref{lig}): $\hat{\Theta}\left( x^{\beta}|\phi^{\nu}\right)$, can be represented according to (\ref{line}), as a Fourier expansion in the momentum-space:
\begin{eqnarray}
\hat{\Theta}\left(x^{\beta}|\phi^{\nu}\right) &=& \frac{1}{(2\pi)^4} \int d^4k \int d^4 s   \nonumber \\
&\times & \left[ A_{s,k}\, e^{i \underleftrightarrow{K}.\overleftrightarrow{X}} e^{\frac{i}{\hbar} \underleftrightarrow{S} \overleftrightarrow{\Phi}}
+ B^{\dagger}_{k,s} \, e^{-i \underleftrightarrow{K}.\overleftrightarrow{X}} e^{-\frac{i}{\hbar} \underleftrightarrow{S} \overleftrightarrow{\Phi}}\right]. \nonumber
\end{eqnarray}
We can define the spinor (complex) components $\hat{\Psi}_{\alpha}\left(x^{\beta}|\phi^{\nu}\right)$
\begin{eqnarray}
 \hat{\Psi}_{\alpha}\left(x^{\beta}|\phi^{\nu}\right)&=& \frac{i}{\hbar (2\pi)^4} \int d^4k \int d^4s \frac{\delta \left(\underleftrightarrow{S} \overleftrightarrow{\Phi}\right)}{\hat{\delta\Phi}^{\alpha}} \left[ A_{s,k}\, e^{i \underleftrightarrow{K}.\overleftrightarrow{X}}  e^{\frac{i}{\hbar} \underleftrightarrow{S} \overleftrightarrow{\Phi}} \right.\nonumber \\
&-&\left. B^{\dagger}_{k,s} \, e^{-i \underleftrightarrow{K}.\overleftrightarrow{X}} e^{-\frac{i}{\hbar} \underleftrightarrow{S} \overleftrightarrow{\Phi}}\right], \nonumber
\end{eqnarray}
where $\left< B \left| A_{ks} B^{\dagger}_{k's'} \right|B\right> = \left(\frac{c^3 M^3_p}{\hbar}\right)^2 \delta^{(4)}(k-k')\,\, \delta^{(4)}(s-s')$, and
\begin{equation}
\frac{\delta }{\hat{\delta\Phi}^{\alpha}}\left(\underleftrightarrow{S} \overleftrightarrow{\Phi}\right) =  \left(2 g_{\alpha\beta}  \mathbb{I}_{4\times 4} - \bar{\gamma}_{\alpha} \bar{\gamma}_{\beta} \right) \hat{S}^{\beta} = 2 \hat{S}_{\alpha} - \bar{\gamma}_{\alpha} \,s=\hat{S}_{\alpha},
\end{equation}
where $s\,\mathbb{I}_{4\times 4}=\frac{1}{4} \,\hat{S}_{\beta} \bar{\gamma}^{\beta}$. Here, $c$ is the speed of light, $M_p$ is the Planckian mass, $\hbar=h/(2\pi)$, $h$-being the Planck constant. Additionally, the squared bi-vectorial $\hat{S}$-norm, is
\begin{equation}
  \left\|\hat{S} \right\|^2 =  \left<B\left|\underleftrightarrow{S} \overleftrightarrow{S}\right|B\right> = \left<B\left| \left( \hat{S}_{\mu} \hat{S}_{\nu} \right) \left( \bar{\gamma}^{\mu} \bar{\gamma}^{\nu}\right) \right|B\right> =s^2
\mathbb{I}_{4\times 4},
\end{equation}
for $\hat{S}_{\mu} = s \,\bar{\gamma}_{\mu}$. In order to quantize the spin, we shall consider the universal invariant
\begin{equation}\label{invariant}
\left<B\left| \underleftrightarrow{S} \overleftrightarrow{\Phi}\right|B\right> = \left<B\left|\left( \hat{S}_{\mu} \hat{\Phi}_{\nu} \right) \left( \bar{\gamma}^{\mu} \bar{\gamma}^{\nu}\right)\right|B\right> =s \phi \,
\mathbb{I}_{4\times 4} = (2\pi n \hbar) \,\mathbb{I}_{4\times 4},
\end{equation}
with $n$-integer. For this reason, gravitons (which have $s=2\hbar$), will be invariant under $\phi=n\,\pi$ rotations and vectorial bosons (with $s =\hbar$), will be invariant under $\phi= 2n\pi$) rotations.

\section{DYNAMICS OF NEUTRAL BOSONS}\label{nb}

Explicitly written, the  dynamics of massless neutral vector bosons is given by\cite{MM}
\begin{eqnarray}
\Box {\hat{\Psi}}^{\alpha}-\nabla_{\beta} \left(\nabla^{\alpha} \hat{\Psi}^{\beta} \right) +  &2& \left( \nabla_{\beta} \hat{\Psi}^{\alpha}\right) \hat{\Psi}^{\beta} - 2 \left(
\nabla_{\beta} \hat{\Psi}^{\beta} \right) \hat{\Psi}^{\alpha} \nonumber \\
- \left(\nabla^{\alpha}\hat{\Psi}^{\gamma}\right) \hat{\Psi}_{\gamma} + &2& \hat{\Psi}^{\alpha} \left(
\nabla_{\gamma} \hat{\Psi}^{\gamma} \right) + \left(\nabla^{\gamma}\hat{\Psi}^{\alpha}\right)\hat{\Psi}_{\gamma}   \nonumber \\
-2 \hat{\Psi}^{\gamma} \left(\nabla_{\gamma} \hat{\Psi}^{\alpha}\right) &=&  2 \left[ \hat{\Psi}^{\mu},
\hat{\Psi}^{\alpha} \right] \hat{\Psi}_{\mu} ,  \label{a2}
\end{eqnarray}
such that in the case of bosons, we obtain
\begin{equation}
 \left< B\left| \left[\hat{\Psi}_{\mu}({\bf x}, {\bf \phi}), \hat{\Psi}_{\nu}({\bf x}', {\bf \phi}') \right]\right|B \right>
 = \frac{s^2}{2 \hbar^2 } \left[\bar\gamma_{\mu} , \bar\gamma_{\nu}\right] \, \sqrt{\frac{\eta}{g}}\,\delta^{(4)} \left({\bf x} - {\bf x}'\right) \,\delta^{(4)} \left({\bf \phi} - {\bf \phi}'\right), \label{coo}
\end{equation}
where $L_p$ is the Planckian length and $\sqrt{\frac{\eta}{g}}$ is the squared root of the ratio between the determinant of the Minkowsky metric: $\eta_{\mu\nu}$ and the metric that describes the background: $g_{\mu\nu}$. This ratio describes the inverse of the relative volume of the background manifold with respect to the Minkowsky one. The Fourier expansion for the spinor field $ \hat{\Psi}_{\alpha}$ is
\begin{eqnarray}\label{fou}
 \hat{\Psi}_{\alpha}&=& \frac{i}{\hbar (2\pi)^4} \int d^4k \int d^4s \frac{\delta \left(\underleftrightarrow{S} \overleftrightarrow{\Phi}\right)}{\hat{\delta\Phi}^{\alpha}} \left[ A_{s,k}\, e^{i \underleftrightarrow{K}.\overleftrightarrow{X}} e^{\frac{i}{\hbar} \underleftrightarrow{S} \overleftrightarrow{\Phi}} \right.\nonumber \\
&-&\left. B^{\dagger}_{k,s} \, e^{-i \underleftrightarrow{K}.\overleftrightarrow{X}} e^{-\frac{i}{\hbar} \underleftrightarrow{S} \overleftrightarrow{\Phi}}\right],
\end{eqnarray}
where $\frac{\delta \left(\underleftrightarrow{S} \overleftrightarrow{\Phi}\right)}{\hat{\delta\Phi}^{\alpha}}=\hat{S}_{\alpha}$. If we deal with bosons, creation and destruction operators must comply\cite{MM}
\begin{eqnarray}
\frac{4 s^2\,   L^2_p}{\hbar^2} \left(|A_{k,s}|^2 - |B_{k,s}|^2\right) &=& 0, \,\pm \left(\frac{c^3 M^3_p}{\hbar}\right)^2. \label{q2}
\end{eqnarray}
The conditions (\ref{q2}) are required for scalar bosons (the first equality) and vector, or
tensor bosons (the second equality). On the other hand, in order for the expectation value of the energy to be positive: $\left< B\left| {\cal H}\right|B \right> \geq 0$, we must choose the negative signature in the second equality of (\ref{q2}). The expectation value for the local particle-number operator for bosons with wave-number norm $k$ and spin $s$, $\hat{N}_{k,s}$, is given by\footnote{To connect the Fock-space theory and the ordinary quantum mechanics one can introduce the wave function in position space by using the definition of a kind of $n_{k,s}$-particle state vector that describes a system of $n_{k,s}$ particles that are localized in coordinate space at the points ${\bf x}_1; {\bf \phi}_1...{\bf x}_n; {\bf \phi}_n$:
\begin{displaymath}
\left|{\bf x}_1,{\bf x}_2,...,{\bf x}_n;{\bf \phi}_1,{\bf \phi}_2,...,{\bf \phi}_n \right> = \frac{1}{\sqrt{n_{k,s}!}} \hat{\slashed{\Psi}}^{\dagger}({\bf x}_1; {\bf \phi}_1)...\hat{\slashed{\Psi}}^{\dagger}({\bf x}_n; {\bf \phi}_n)\left|B\right>,
\end{displaymath}
where here $\left|B\right>$ is our reference state. This state is not a vacuum state because it describes a curved background state, but describes the Riemannian (classical) reference with respect to which we describe the quantum system.}
\begin{equation}\label{np}
\left<B\left|\hat{N}_{k,s}\right|B\right> =-n_{k,s}\,\left(\frac{\hbar}{c^3 M^3_p}\right)^2\,\int d^4x \sqrt{-g} \int d^4\phi \,\left< B\left| \left[\hat{\slashed{\Psi}}({\bf x}, {\bf \phi}), \hat{\slashed{\Psi}}^{\dagger}({\bf x}, {\bf \phi}) \right]\right|B \right>=n_{k,s} \,\mathbb{I}_{4\times4},
\end{equation}
where the slashed spinor fields are: $\hat{\slashed{\Psi}}=\bar{\gamma}^{\mu} {\hat\Psi}_{\mu}$, $\hat{\slashed{\Psi}}^{\dagger} = \left(\bar{\gamma}^{\mu} {\hat\Psi}_{\mu}\right)^{\dagger}$. Furthermore, these fields comply with the algebra
\begin{equation}
\left< B\left| \left[\hat{\slashed{\Psi}}({\bf x}, {\bf \phi}), \hat{\slashed{\Psi}}^{\dagger}({\bf x}', {\bf \phi}') \right]\right|B \right>
 = \frac{4 s^2 \,L^2_p}{\hbar^2} \left(|A_{k,s}|^2 - |B_{k,s}|^2\right) \,\,  \sqrt{\frac{\eta}{g}} \,\,\delta^{(4)} \left({\bf x} - {\bf x}'\right) \,\delta^{(4)} \left({\bf \phi} - {\bf \phi}'\right),
\end{equation}
which must be nonzero in order to particles can be created. Notice that this is the case for bosons with spin non-zero, but in the case of scalar bosons, which have zero spin, one obtains that $\left(|A_{k,s}|^2 - |B_{k,s}|^2\right)=0$, and $\left< B\left| \left[\hat{\slashed{\Psi}}({\bf x}, {\bf \phi}), \hat{\slashed{\Psi}}^{\dagger}({\bf x}', {\bf \phi}') \right]\right|B \right>=0$. This result is valid in any relativistic scenario.

If we take the expectation value for (\ref{a2}), and we take into account (\ref{coo}) and (\ref{fou}), we obtain the following equation for the wave-numbers of bosons:
\begin{eqnarray}
\left[\bar{\gamma}^{\beta},\, \bar{\gamma}^{\theta}\right]_{,\theta} &-& \frac{1}{2} g^{\beta\theta} \left(\bar{\gamma}^{\nu}\right)_{,\theta} \bar{\gamma}_{\nu}
- 2i\, k^{\beta} \mathbb{I}_{4\times4} + \frac{1}{2} g^{\nu\theta} \left(\bar{\gamma}^{\beta}\right)_{,\theta} \bar{\gamma}_{\nu} +\frac{i}{2} \bar{\gamma}^{\beta} \underleftrightarrow{k} \nonumber \\
&= &  \frac{s^2}{2\hbar^2} \left\{ \begin{array}{cc}  \nu \, \\ \theta \, \nu  \end{array} \right\} \left[
\bar{\gamma}^{\theta}, \, \bar{\gamma}^{\beta} \right] + \frac{1}{2} g^{\beta\theta}  \left\{ \begin{array}{cc}  \mu \, \\ \nu \, \theta  \end{array} \right\} \bar{\gamma}^{\nu}
\bar{\gamma}_{\mu} - \frac{1}{2} g^{\mu \theta}  \left\{ \begin{array}{cc}  \beta \, \\ \nu \, \theta  \end{array} \right\} \bar{\gamma}^{\nu} \bar{\gamma}_{\mu}, \label{em}
\end{eqnarray}
where $\underleftrightarrow{k} = k^{\alpha} \bar{\gamma}_{\alpha}$ and $\bar{\gamma}_{\alpha}= E^{\mu}_{\alpha} \gamma_{\mu}$ are the components of the basis on the background metric, which are related by the vielbein $E^{\mu}_{\alpha}$ with the $4\times 4$ matrices $\gamma^{\mu}$ on the Minkowsky spacetime. In our case we shall use cartesian coordinates to describe spacial coordinates. In this paper we shall use the Weyl representation of the $\gamma$-matrices to generate the hyperbolic space-time.

\section{Pre-inflation}\label{pin}

The idea of a pre-inflationary expansion of the universe in which the universe begins to expand through a (global) topological phase transition was proposed in\cite{mb1}.
In this model was studied the birth of the universe using a complex time $\tau(t)=\int e^{i \hat{\theta}(t)}dt$, such that the phase transition from a pre-inflationary to inflationary
epoch was examined using a dynamical rotation of the complex time, $\tau(t)$, on the complex plane. After a particular choice of coordinates, one can define a dynamical variable $\theta$:
$\pi/2 \geq \hat{\theta}(t)>0$, such that it describes the dynamics of the system and it is related with the expansion of the universe
\begin{equation}
\hat\theta(t)=\frac{\pi}{2} e^{-H_0 t}.
\end{equation}
We consider the line element introduced in\cite{mb1} to describe pre-inflation
\begin{equation}\label{m}
d\hat{S}^2 =  \left(\frac{\pi a_0}{2}\right)^2 \frac{1}{\hat{\theta}^2} \left[{d\hat{\theta}}^2 - \delta_{ij} d\hat{x}^i d\hat{x}^j\right],
\end{equation}
If we desire to describe an initially Euclidean 4D universe, that thereafter evolves to an asymptotic value $\hat{\theta} {\rightarrow } 0$, we must require that $\hat{\theta}$ to have an initial value $\hat\theta_0=\frac{\pi}{2}$. Furthermore, the nonzero components of the Einstein tensor, are
\begin{equation}
G_{00} = - \frac{3}{\hat{\theta}^2} , \qquad G_{ij} =  \frac{3}{\hat\theta^2 } \,\delta{ij},
\end{equation}
so that the energy density and the pressure, are respectively given by
\begin{equation}
\rho(\hat\theta) = \frac{1}{\pi G} \frac{3}{(\pi a_0)^2}, \qquad\qquad
P(\hat\theta) = - \frac{1}{\pi G} \frac{3}{(\pi a_0)^2}.
\end{equation}
The equation of state for the metric (\ref{m}), is
\begin{equation}
\frac{P}{\rho} =  - 1.
\end{equation}

We shall describe the case where the asymptotic evolution of the Universe is described by a vacuum expansion. In this case the asymptotic scale factor, Hubble parameter and the
potential are are respectively given by
\begin{equation}
a(t)= a_0\, e^{H_0 t}, \qquad \frac{\dot{a}}{a} = H_0 \qquad V= \frac{3}{8\pi G} H^2_0,
\end{equation}
so that, due to the fact that $\frac{\delta V}{\delta\phi}=0$, the background field background solution of the background dynamics
\begin{equation}
\phi{''}-\frac{2}{\hat\theta} \phi{'}=0,
\end{equation}
is
\begin{equation}
\phi(t)= \phi_0.
\end{equation}
This solution describes the background solution of the field that drives a phase transition of the global geometry
from a 4D Euclidean space to a 4D hyperbolic spacetime. The exact back-reaction effects were considered in \cite{rb}. In the present paper we shall consider
the emission of gravitons (massless bosons of spin $2$ and spin $1$ using unified spinor fields\cite{MM}.

\subsection{Graviton's emission in pre-inflation}

To study the graviton's emission during pre-inflation we shall use the equation (\ref{em}), for spin $s=2\hbar$ [see eq. (\ref{coo})]. In this case the eq. (\ref{em}) can be separated in two new equations
\begin{eqnarray}
\left[ \bar{\gamma}^{i},\bar{\gamma}^0\right]_{,0} &= & \frac{s^2}{8 \hbar^2}  \left\{ \begin{array}{cc}  \nu \, \\ 0 \, \nu  \end{array} \right\} \left[\bar{\gamma}^0, \bar{\gamma}^{i}\right], \label{g1}\\
k^{\beta} \mathbb{I}_{4\times4} & = & \frac{1}{6} k^{\alpha} \left[\gamma^{\beta}, \gamma_{\alpha} \right] + \frac{3 i s^2 }{8 \hbar^2}  \left\{ \begin{array}{cc}  \nu \, \\ \theta \, \nu  \end{array} \right\} \left[ \bar{\gamma}^{\theta}, \bar{\gamma}^{\beta} \right]. \label{g2}
\end{eqnarray}
The equations (\ref{g1}) and (\ref{g2}) give us the conditions that must be fulfilled by the modes with wave-number $k$ and massless bosons with spin $s=2\hbar$, in the equation (\ref{a2}). In particular, the equation (\ref{g1}) is fulfilled only by $s=2\hbar$-spin massless bosons (gravitons). During pre-inflation $\bar{\gamma}^{\beta}=\frac{2\hat{\theta}}{\pi a_0} \gamma^{\beta}$ and $\left(\bar{\gamma}^{\beta}\right)_{,0} = \frac{1}{\hat{\theta}} \bar{\gamma}^{\beta}$. From (\ref{g2}) we obtain four vector equations that provide us the solutions for the $k^{\beta}$-components of the graviton's propagation during pre-inflation. Using the fact that (for $i,j=1,2,3$)
\begin{equation}
\left\{ \begin{array}{cc}  0 \, \\ 0 \, 0  \end{array} \right\}=\left\{ \begin{array}{cc}  1 \, \\ 0 \, 1  \end{array} \right\}=\left\{ \begin{array}{cc}  2 \, \\ 0 \, 2  \end{array} \right\}=\left\{ \begin{array}{cc}  3 \, \\ 0 \, 3  \end{array} \right\} =-\frac{1}{\hat{\theta}},
\end{equation}
we obtain the resulting values for $k^{\beta}$
\begin{equation}
k^0= -\left(\frac{18\,i}{\pi a_0}\right)\,\hat{\theta}, \qquad k^1=  k^2=0, \qquad k^3= \pm  \,\left(\frac{74\,i}{\pi a_0}\right)\,\hat{\theta},
\end{equation}
such that the physical wavenumber-norm of gravitons that propagate in the $\hat{z}$-direction, is
\begin{equation}
\frac{|k|^2}{a^2(\hat{\theta})}=\frac{\left(k_{\alpha} k^{\alpha}\right)}{a^2(\hat{\theta})}=\left[\frac{1288}{\left(\pi a_0\right)^2} \right]\hat{\theta}^2 >0.
\end{equation}
Notice that tends to zero with the expansion of the universe, due to the fact $\hat{\theta} \rightarrow 0$ with the increasing of the scale factor $a(\hat{\theta})=\frac{\pi a_0}{2 \hat{\theta}}$. However, due to the fact we are dealing with photons, the $k$-squared-norm must be null on physical coordinates. Therefore, the effective frequency and the $z$-component of the wave-number on physical coordinates should be altered in the following manner (we use natural units):
\begin{equation}
\omega^2 \equiv \left(\tilde{k}^0\right)^2 = \left[\Im\left(\frac{{k}^0}{a(\hat{\theta})}\right)\right]^2 + \frac{\left|k\right|^2}{a^2}=
\left(\frac{74}{\pi a_0}\right)^2\,\hat{\theta}^2, \qquad \left(\tilde{k}^3\right)^2 =\left[\Im\left(\frac{{k^3}}{a(\hat{\theta})}\right)\right]^2,
\end{equation}
such that $\tilde{k}_{\alpha}\tilde{k}^{\alpha}=0$. Therefore, the redefined physical values $\tilde{k}^{\alpha}$, should be the values experimentally measured. The physical wavelength results to be $\lambda_{ph}=\left(\frac{2\pi}{37}\right) \left(\frac{a(\hat{\theta})}{a_0}\right) H^{-1}_0$, such that $H_0=a^{-1}_0$. In other words the physical wavelength of gravitons is something smaller than the physical Hubble radius during pre-inflation.

\subsection{Massless $s=\hbar$-bosons emission in pre-inflation}

In order to study the massless $s=\hbar$ bosons emitted during the pre-inflationary epoch, we shall use the equation (\ref{em}), with (\ref{coo}), for $s=\hbar$. In this
case the equation (\ref{em}) can be splited in two equations
\begin{eqnarray}
\left[ \bar{\gamma}^{i},\bar{\gamma}^0\right]_{,0} &= & \frac{s^2}{2 \hbar^2}  \left\{ \begin{array}{cc}  \nu \, \\ 0 \, \nu  \end{array} \right\} \left[\bar{\gamma}^0, \bar{\gamma}^{i}\right], \label{f1}\\
k^{\beta} \mathbb{I}_{4\times4} & = & \frac{1}{2} k^{\alpha} \left[\gamma^{\beta}, \gamma_{\alpha} \right]. \label{f2}
\end{eqnarray}
The equation (\ref{f1}) is needed to assure its validity for $s=\hbar$, so that the equation (\ref{f2}) is fulfilled in order to obtain the wavenumber components of the wave. Notice that
if we sum the equations (\ref{f1}) and (\ref{f2}), we obtain the equation (\ref{em}) for the metric (\ref{line}). The wave-number solutions for the four equations (\ref{f2}), are
\begin{equation}
k^0= -\left(\frac{i}{\pi^2 a^2_0}\right)\,\hat{\theta}, \qquad k^1=  k^2=0, \qquad k^3= \mp  \,\left(\frac{3\,i}{\pi a_0}\right)\,\hat{\theta}.
\end{equation}
The physical wavenumber-norm of $s=\hbar$-bosons that propagate in the $\hat{z}$-direction, is:
\begin{equation}
\frac{|k|^2}{a^2(\hat{\theta})}=\frac{\left(k_{\alpha} k^{\alpha}\right)}{a^2(\hat{\theta})}=\left[\frac{8}{\left(\pi a_0\right)^2} \right]\hat{\theta}^2 >0.
\end{equation}
which, as in the case of gravitons, tends to zero with the expansion of the universe. In order for the $k$-squared-norm be null, the frequency and $z$-wave number must be altered on physical coordinates, $\tilde{k}_{\alpha}\tilde{k}^{\alpha}=0$:
\begin{equation}
\omega^2 \equiv \left(\tilde{k}^0\right)^2 = \left[\Im\left(\frac{{k}^0}{a(\hat{\theta})}\right)\right]^2 + \frac{\left|k\right|^2}{a^2}=
\left(\frac{3}{\pi a_0}\right)^2\,\hat{\theta}^2,\qquad \left(\tilde{k}^3\right)^2 =\left[\Im\left(\frac{{k^3}}{a(\hat{\theta})}\right)\right]^2.
\end{equation}
These should be the values measured in an experiment. In this case, the physical wavelength for massless $s=\hbar$-bosons, is: $\lambda_{ph}=\left(\frac{4\pi}{3}\right) \left(\frac{a(\hat{\theta})}{a_0}\right) H^{-1}_0$, which is something bigger than both, the physical graviton's wavelength and than the Hubble horizon.

\section{Final Comments}\label{fc}

Following the unified spinor field theory recently introduced, we
have obtained the universal equation of motion for massless bosons
with quantization included: (\ref{em}). This equation describes
the dynamics of massless bosons with different spin on arbitrary
Riemannian background. The dynamics of some particular spinor
field is given when we consider the equation (\ref{coo}) in the
background (Riemannian) expectation value of the equation
(\ref{a2}). In particular, we have explored the case of a
pre-inflationary scenario described in Sect. \ref{pin}. In this
epoch the universe suffered a global topological phase transition
that made possible the transition between a global 4D Euclidean
universe and an hyperbolic one through an expansion governed by a
vacuum equation of state: $P=-\rho$. A remarkable result here
obtained is that during this epoch gravitons and $s=\hbar$-bosons
take a positive relativistic squared norm in physical coordinates:
$\left.\frac{|k|^2}{a^2(\hat{\theta})}\right|_{s=(1,2)\hbar} >0$,
which tends to zero with the expansion of the universe. However,
by redefining the physical coordinates in order to obtain
$\tilde{k}_{\alpha} \tilde{k}^{\alpha}=0$, we obtain that the
wavelengths of both, gravitons and photons is increase co-moving
with the Hubble radius of the universe: $\lambda_{Ph} \sim a/H_0$.
Our calculations show that gravitational radiation is emitted
during the big bang. As was shown in (\ref{np}), bosons with
$s=(1,2)\hbar$ can be created in any relativistic scenario, and
therefore can be created during pre-inflation. However, this is
not the case of scalar bosons, which have zero spin. Because the
wave-length of both, photons and gravitons are of the order of the
Hubble horizon, the frequency should be very low, of the order of
the inverse of the edge of the universe, which make it very
difficult to be detected because the diluting effects of the
expansion of the universe. However, they should be responsible for
very large-scale gravitational en electromagnetic primordial
fundamental wavelengths, which are coherent, and could be detected
in the future on the extreme (low-frequencies) range of the
primordial electromagnetic and gravitational spectrum.

\section*{Acknowledgements}

\noindent The authors acknowledge CONICET, Argentina (PIP 11220150100072CO) and UNMdP (EXA852/18), for financial support.


\begin{thebibliography}{99}
\bibitem{hei1} W. Heisenberg, Rev. Mod. Phys. {\bf 29}, 269 (1957).
\bibitem{hei2} W. Heisenberg, Naturwiss. {\bf 61}, 1 (1974).
\bibitem{hohuang} Tin-Lun Ho and Biao Huang, Phys. Rev. Lett. {\bf 115}, 155304 (2015).
\bibitem{weyl} H. Weyl, {\em Philosophy of Mathematics and Natural Science}, english version, Princeton University Press (1949).
\bibitem{aa} A. Ashtekar, J. Lewandowski, Class. Quant. Grav. {\bf 21}: R53–R152 (2004); \\
C. Rovelli, Living Rev. Rel. {\bf 1}: 1 (1998).
\bibitem{MM} M. R. A. Arcod\'{\i}a, M. Bellini, {\em Towards unified spinor fields: confinement of gravitons on a dS background}. E-print: arXiv 1703.01355.
\bibitem{rb} L. S. Ridao, M. Bellini, Phys. Lett. {\bf B751}, 565 (2015); \\
L. S. Ridao, M. Bellini, Astrophys. Space Sci., {\bf 357}, 94 (2015).
\bibitem{mb1} M. Bellini, Phys. Lett. {\bf B771}: 227-229 (2017).
\bibitem{la} K. Lazaridis {\em et al}, Mon. Not. Roy. Astron. Soc. {\bf 414}: 3134 (2011).
\end{thebibliography}
\end{document}